\documentclass[prl,twocolumn,groupedaddress,aps]{revtex4-1}

\def\myfontsize{}
\def\ket#1{\left|#1\right\rangle}
\def\trb{{\rm Tr}_{\rm B}\mathop}
\makeatletter
\def\Hy@safe@activestrue{}
\makeatother

 \advance\textfloatsep by -0.1in
 \advance\abovecaptionskip by -0.05in
 \advance\belowcaptionskip by -0.06in

\usepackage{graphicx}
\begin{document}

\title{Universal set of scalable dynamically corrected gates for quantum error
  correction with always-on qubit couplings}

\author{Amrit De}
 \affiliation{Department of Physics \& Astronomy, University of California, Riverside, California 92521}
\author{Leonid P. Pryadko}
 \affiliation{Department of Physics \& Astronomy, University of California, Riverside, California 92521}
\date{\today}

\begin{abstract}
  We construct a universal set of high fidelity quantum gates to be used on a
  sparse bipartite lattice with always-on Ising couplings.  The gates are
  based on dynamical decoupling sequences using shaped pulses, they protect
  against low-frequency phase noise, and can be run in parallel on
  non-neighboring qubits.  This makes them suitable for implementing quantum
  error correction with low-density parity check codes like the surface codes
  and their finite-rate generalizations.  We illustrate the construction by
  simulating quantum Zeno effect with the $[[4,2,2]]$ toric code on a spin
  chain.
\end{abstract}

\maketitle

Quantum error correction (QEC) makes it theoretically possible to perform
large quantum computations with a finite per-qubit error
rate\cite{shor-error-correct,Knill-Laflamme-1997,Bennett-1996}.  In practice
QEC is extremely difficult since the corresponding \emph{error probability
  threshold} is small\cite{Knill-error-bound,%Rahn-2002,
  Dennis-Kitaev-Landahl-Preskill-2002,%
  Steane-2003,%
  Fowler-QEC-2004,Fowler-2005,%fowler-thesis-2005,%
  %Knill-nature-2005,Knill-2005,
  Raussendorf-Harrington-2007}.  When only local interactions between the qubits are allowed, the estimated threshold value is the highest,
around $1\%$, for toric and related surface
codes\cite{kitaev-anyons,Dennis-Kitaev-Landahl-Preskill-2002,%
  Raussendorf-Harrington-2007}.
However, as to how to implement the operations efficiently and with the required accuracy, is an open question.
%%% Question is, how to implement the
%%% operations efficiently and with required accuracy.

Qubits with always-on couplings are a natural model for several potential quantum computer (QC) architectures currently under investigation.
%[nmr,Josephson,P in Si,e on He,...].
On general grounds, compared to their counterparts with tunable couplings, such qubits can be expected to have better parameter
stability and longer coherence times.  In addition, over sixty years of development in nuclear magnetic resonance (NMR) yielded an amazing
degree of control available to such systems\cite{vandersypen-2004,Criger-2012}.  Related techniques based
on selective dynamical decoupling(DD) of parts of the system Hamiltonian with carefully designed pulse sequences have been further developed in
application to quantum computation\cite{viola-knill-lloyd-1999B,jones-1999,%
  Viola-2002,Viola-Knill-2003,khodjasteh-Lidar-2005,Uhrig-2007,Souza-2011}.

While NMR quantum computation is not scalable\cite{jones-2001}, it still holds
several records for a number of coherently controlled
qubits\cite{Criger-2012}.  However, some of these have been achieved with the
help of the \emph{strongly-modulated} pulses, computer-generated single- and
multi-qubit gates tailored for a particular system
Hamiltonian\cite{price-1999,price-1999-jmr,price-havel-cory-2000,%
  Fortunato-Cory-smp-2002}.
While such gates can be used in other QC architectures\cite{Vartiainen-2004},
they may violate the scalability.

On the other hand, NMR-inspired techniques like DD
can also be used to control large systems with local interactions, with the
scalability achieved by designing pulses and sequences to a given order in the
Magnus series\cite{slichter-book} on small qubit
clusters\cite{sengupta-pryadko-ref-2005,%
  pryadko-sengupta-kinetics-2006}.  DD is also excellent in producing accurate
control for systems where not all interactions are known as one can decouple
interactions with the given
symmetry\cite{stollsteimer-mahler-2001,Tomita-2010}.  Moreover, DD works best
against errors coming from low-frequency bath degrees of freedom which tend to
dominate the decoherence rates, and it does not require additional qubits.  In
short, DD is an excellent choice for the first level of
coherence protection; it's use could greatly reduce the required repetition
rate of the QEC cycle.

This is well recognized in the community and applications of DD for QC
are actively investigated by a number of groups.  However, most
publications on the subject illustrate general principles using a
single qubit as an example, leaving out the issues of design and
simulation of scalable approaches to multi-qubit dynamical decoupling.
While the techniques for larger systems exist, they typically yield
longer decoupling
sequences\cite{stollsteimer-mahler-2001,%
  khodjasteh-Lidar-2005,Khodjasteh-Viola-PRL-2009}.

\textsc{The goal of this work} is to provide a scalable benchmark
implementation of a universal set of accurate gates using soft pulses for a
system with always-on qubit couplings.  Specifically, we construct one- and
two-qubit gates with built-in DD-protection against low-frequency phase noise
for a sparse bipartite lattice of qubits with the nearest-neighbor (n.n.)\
Ising couplings.  The constructed gates use finite-amplitude \emph{shaped
  pulses} which can be implemented experimentally. They are \emph{scalable}, in the
sense that the same construction works for an arbitrary lattice, and they can
also be executed \emph{in parallel} for different qubits and/or qubit pairs.
This makes them ideal for implementing QEC with quantum low-density
parity check (LDPC) codes\cite{Postol-2001,MacKay-Mitchison-McFadden-2004}, in
particular, the surface codes and their finite-rate
generalizations\cite{Dennis-Kitaev-Landahl-Preskill-2002,%
  Tillich-Zemor-2009,Kovalev-Pryadko-2012}.  In the limit of very slow
(classical) bath the gates are accurate to second order in the Magnus
expansion, meaning that their infidelity scales as sixth or higher powers of
the coupling, in units of inverse pulse duration.  We demonstrate the accuracy
of the constructed gates by simulating the quantum Zeno
effect\cite{Facchi-Pascazio-2002,Facchi-2002B} repeatedly for the $[[4,2,2]]$
error-detecting toric code on an Ising chain. The simulations are done with
five qubits, using classical correlated noise as a source of
dephasing.

Two techniques are essential to our work.  First, the use of NMR-style
self-refocusing pulses\cite{warren-herm,sengupta-pryadko-ref-2005,%
  pryadko-sengupta-2008,Pasini-2008}, which (to a given order) work as
drop-in replacement for hard, $\delta$-function-like pulses.  In
simulations, we use the second-order  pulses designed
and characterized in Refs.~\onlinecite{sengupta-pryadko-ref-2005,%
  pryadko-sengupta-kinetics-2006,%
  pryadko-quiroz-2007,pryadko-sengupta-2008}.  The second technique is
the Eulerian path construction for generating accurate DD
sequences\cite{Viola-Knill-2003}, and its extension, the dynamically
corrected gates\cite{Khodjasteh-Viola-PRL-2009,%
  Khodjasteh-Viola-PRA-2009,Khodjasteh-Lidar-Viola-2009} which allow
for the construction of composite pulses accurate to a given order of
the Magnus expansion.

\textsc{We construct our gates} for a collection of qubits arranged on an arbitrary
sparse bipartite graph $\mathcal{G}$ with edge set $\mathcal{E}$, with an
Ising coupling for every edge,
\begin{equation}
  \label{eq:system}
  H_{\rm S}\equiv {1\over 2} \sum_{(ij)\in \mathcal{E}} J_{ij}  \sigma^z_i\sigma^z_j,
\end{equation}
arbitrary (within the bandwidth) single-qubit control,
\begin{equation}
  \label{eq:control}
  H_{\rm C}\equiv {1\over 2} \sum_{i}\sum_{\mu=x,y,z} \sigma^\mu_iV_{i\mu}(t),
\end{equation}
in the presence of low-frequency phase noise
\begin{equation}
  \label{eq:noise}
  H_{\rm N}\equiv {1\over 2} \sum_{i} \sigma^z_i B_{i}+H_{\rm B},
\end{equation}
where the bath coupling operators $B_i$ can result, e.g., from low-frequency
phonons, or nuclear spins, and their dynamics is governed by the bath
Hamiltonian $H_{\rm B}$ independent from $\sigma_i^\mu$.

Decoherence resulting from higher-frequency bath modes, e.g., described by the
Lindblad equation\cite{lindblad-76}, can be also introduced, but at later
design stages, since DD is not effective against such decoherence.  While we
do not consider Markovian decoherence here, we mention in passing that the
main effects of DD are the suppression of equilibrium population asymmetries
(qubits are constantly flipped), and, with soft-pulse DD, the redistribution
of decoherence rates between the channels\cite{pryadko-quiroz-2007}.  For
example, even if dephasing is dominant for non-driven qubits, any sequence of
finite-width pulses creates some longitudinal relaxation (compensated by a
reduction of the dephasing rate).

To construct
the CNOT gate, we use the identity
%\cite{[{See,~for~example, Eq.~(6) in }]Galiautdinov-Geller-2007}:
\cite{Galiautdinov-Geller-2007}
%%% Exp[I Pi/4]*
%%% MatrixExp[-I Pi/4 Sy**S0].
%%% MatrixExp[-I Pi/4 S0**Sx].
%%% MatrixExp[I Pi/4 Sx**S0].
%%% MatrixExp[I Pi/4 Sy**S0].
%%% MatrixExp[I Pi/4 S0**Sy].
%%% MatrixExp[-I Pi/4 Sz**Sz].
%%% MatrixExp[-I Pi/4 S0**Sy]
\begin{equation}
\mathtt{CNOT}_{cd}=e^{i\pi/4}Y_c\bar X_c\bar Y_c X_d\bar Y_d e^{-i\pi/4
  \sigma^z_c\sigma^z_d} Y_d,
\label{eq:cnot-decomposition}
\end{equation}
where $X_i\equiv \exp \left(-i {\pi\over 4}\sigma^x_i\right)$, $Y_i\equiv
\exp\left( -i {\pi\over 4}\sigma^y_i\right)$, are $\pi/2$ unitaries, and $\bar
X_i$, $\bar Y_i$ denote the corresponding conjugate gates [in simulations we
use the equivalent form with $Y_c\bar X_c\bar Y_c\equiv Z_c$].

\begin{figure}[htbp]
  \center
  \includegraphics[width=0.8\columnwidth]{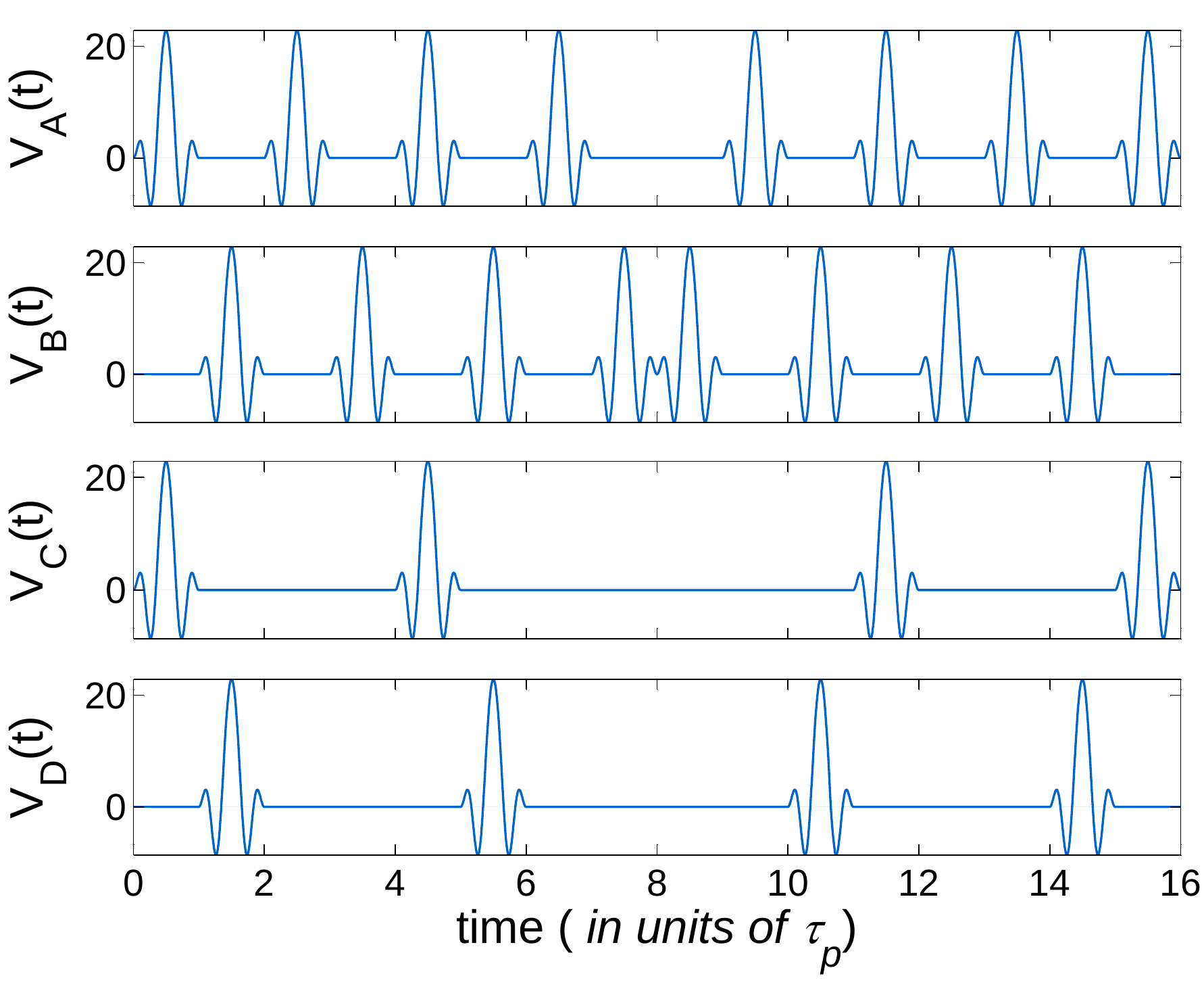}
  \caption{\myfontsize (color online) Pulse sequences used to implement
    two-qubit $zz$-rotations on a bipartite lattice with Ising couplings.
    Global sequences of $\pi$-pulses in $x$ direction $V_A(t)$ and $V_B(t)$
    are executed on the idle qubits of the two sublattices.  These decouple
    the inter-qubit Ising couplings and also the single-qubit low-frequency
    phase noise terms.  In order to couple two neighboring qubits, on these
    qubits the sequences $V_A$, $V_B$ are replaced by $V_C$, $V_D$,
    respectively.  This produces an effective Ising Hamiltonian with half of
    the original coupling.  Shown are $Q_1(\pi)$ second-order self-refocusing
    pulse shapes\cite{sengupta-pryadko-ref-2005,pryadko-sengupta-2008} used in
    the simulations.}
  \label{fig:zz}
\end{figure}

\textsc{To implement the two-qubit} $zz$-rotation gate, $e^{-i\pi/4
  \sigma^z_c\sigma^z_d}$, we run two period-$16\tau_p$ decoupling sequences on
the sublattices $A$ and $B$, $V_A(t)$ and $V_B(t)$ in Fig.~\ref{fig:zz}, where
each pulse of duration $\tau_p$ is a symmetric $\pi$ pulse applied in the $x$
direction.  When the pulses are second-order self-refocusing pulses [e.g.,
$Q_1(\pi)$ from
Refs.~\onlinecite{sengupta-pryadko-ref-2005,pryadko-sengupta-2008} shown],
these sequences suppress the effect of the Ising couplings $H_{\rm S}$ and the
noise $H_{\rm N}$ to second order in the Magnus expansion, meaning that the
effective Hamiltonian is just $H_{\rm B}$, with the error scaling as
$\propto\tau_p^2$.  This gives the error in the unitary matrix scaling as
$\propto\tau_p^3$, and the corresponding infidelity scaling as
$\propto\tau_p^6$.

To turn on the coupling between two neighboring qubits $c$, $d$, the two
sublattice sequences on these qubits are replaced with $V_C(t)$ and $V_D(t)$,
respectively, see Fig.~\ref{fig:zz}.  These sequences are chosen so that the
Ising coupling between these qubits is only removed half of the time, while
the coupling to other qubits continues to be removed.  More precisely, with
second-order pulses, the effective Hamiltonian is $H_{\rm
  B}+(J/4)\sigma^z_c\sigma^z_d+\mathcal{O}(\tau_p^2)$.  Repeating this
sequence $m$ times gives the system evolution
\begin{equation}
  U_{zz}=\exp(-i \alpha
\sigma_c^z\sigma_d^z),\quad \alpha=4m J\tau_p,\label{eq:Uzz}
\end{equation}
with the error scaling as $\propto m\tau_p^3$, where the term associated with
the bath evolution is suppressed.  Such sequences can be run simultaneously on
many pairs of qubits as long as qubits from different pairs are not mutually
coupled.

\begin{figure}[htbp]
  \centering
  \includegraphics[width=0.75\columnwidth]{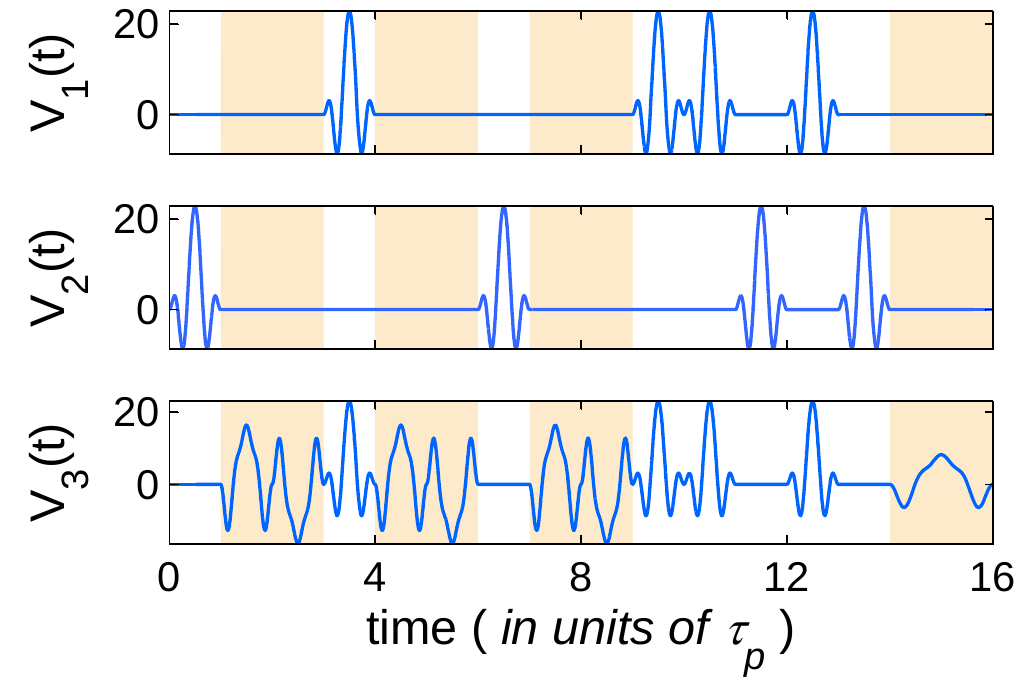}
  \caption{\myfontsize (color online) Pulse sequence used to implement
    single-qubit rotations.  The sequences of $\pi$ pulses in $x$ direction,
    $V_1(t)$ and $V_2(t)$, are executed globally on the idle qubits of the two
    sublattices.  A single-qubit rotation is implemented as a
    DCG\cite{Khodjasteh-Viola-PRL-2009,Khodjasteh-Viola-PRA-2009} by adding
    three pulse-antipulse combinations and the stretched pulse in the shaded
    regions; the sequence $V_3(t)$ where $Q_1(\pi/2)$ pulses in $y$ direction
    are added corresponds to a $(\pi/2)_Y$ operation on a sublattice-A qubit.
    Such gates can be executed on any set of non-neighboring qubits.}
  \label{fig:pulseY3p}
\end{figure}

\textsc{We implement single-qubit rotations} with the leading-order
DCGs\cite{Khodjasteh-Viola-PRL-2009,Khodjasteh-Viola-PRA-2009}, using the
pulse sequences in Fig.~\ref{fig:pulseY3p}.  Again, two decoupling sequences,
$V_1(t)$ and $V_2(t)$, are run globally on the two sublattices; additional
pulses are inserted for the qubits to be rotated [$V_3(t)$ in
Fig.~\ref{fig:pulseY3p} shows an implementation of the $\pi/2$ rotation with
respect to $Y$ axis on a sublattice-A qubit].

Nominally, DCG guarantees the first-order decoupling with any pulses.
However, in our case, the decoupling sequences $V_i(t)$, $i=1,2$, do not go
over the complete single-qubit groups.  Thus, unoptimized (e.g., Gaussian)
pulses can produce unitary errors scaling linearly with the pulse duration
$\tau_p$; one needs first-order self-refocusing
pulses\cite{warren-herm,sengupta-pryadko-ref-2005,pryadko-sengupta-2008} to
get first-order decoupling.  In the case of the second-order pulses (e.g.,
$Q_1(\phi)$ \cite{pryadko-sengupta-2008}), the remaining
order-$\tau_p^2$ errors are all proportional to different commutators
$[B_i,B_j]$ and $[H_{\rm B},B_i]$, which gives second-order decoupling
(infidelity $\propto \tau_p^6$) when the operators $B_i$ are replaced by
$c$-numbers $\Delta_i$ (cf.\ chemical shifts in NMR).

\begin{figure}[htbp]
  \centering
  \includegraphics[width=0.9\columnwidth]{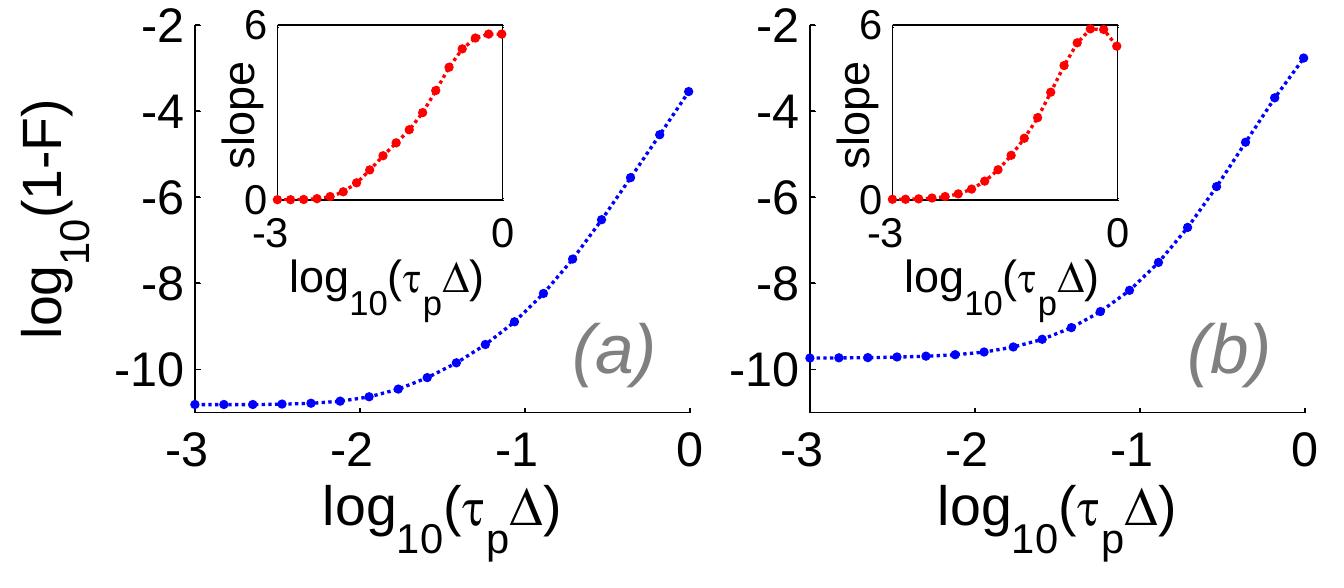}
  \caption{\myfontsize (color online) (a) The infidelity of the gate in
    Fig.~\ref{fig:pulseY3p} and (b) the infidelity of the complete C-{\sc not}
    gate between qubits 2 and 3 of a four-qubit n.n.\ Ising chain as a
    function of r.m.s.\ chemical shift $\Delta$ (pulse sequences not shown).
    For small $\Delta$ the errors saturate because of the fixed interqubit
    coupling $J$; for large $\Delta$ the errors are dominated by the chemical
    shifts.  The insets show the corresponding slopes; the slope of
    approximately $6$ [infidelity $\propto (\tau_p \Delta)^6$] indicate that
    the on-site chemical shifts are decoupled to quadratic order.  Analogous
    calculations with first-order
    pulses\cite{sengupta-pryadko-ref-2005,pryadko-sengupta-2008} $S_1(\pi)$
    and $S_1(\pi/2)$ give 1st-order [infidelity $\propto (\tau_p\Delta)^4$]
    decoupling and two orders of magnitude higher infidelities (not shown);
    Gaussian pulses increase infidelities by up to five orders of magnitude
    depending on the width.}
  \label{fig:F}
\end{figure}

These accuracy predictions are confirmed in Fig.~\ref{fig:F} which shows the
average infidelities for a single $(\pi/2)_Y$ rotation of qubit 3
[Fig.~\ref{fig:F}(a)] and a complete CNOT$_{23}$ gate [Fig.~\ref{fig:F}(b)] as
a function of the r.m.s.\ chemical shift $\Delta$ (in units of $\tau_p^{-1}$),
obtained numerically for a four-qubit Ising chain.  The simulations are done
with a custom {\tt C++} program using fourth-order Runge-Kutta algorithm for
integrating the unitary dynamics and the \texttt{Eigen3}
library\cite{eigenweb} for matrix arithmetics.  We fix the value of
$J_{ij}=J=\pi/(16m\tau_p)$ with $m=5$ repetitions of the basic sequence [see
Eqs.~(\ref{eq:cnot-decomposition}), (\ref{eq:Uzz}) and Fig.~\ref{fig:zz}] in
the CNOT gate; with the addition of four single-qubit DCGs [see
Fig.~\ref{fig:pulseY3p}] the CNOT duration is $t_{\rm CNOT}=9\times
16\tau_p=144\tau_p$.  For small $\Delta$, the infidelities are dominated by
the decoupling accuracy of the inter-qubit interactions, while they scale as
$\propto \Delta^6\tau_p^6$ for large $\Delta$, see the graphs of the
corresponding slopes in the insets.

\textsc{We illustrate the performance} of the designed gates by simulating the
quantum Zeno effect\cite{Facchi-Pascazio-2002,Facchi-2002B} using the
four-qubit toric error-detecting code %$[[4,2,2]]$
\cite{kitaev-anyons,Grassl-Beth-Pellizzari-1997}.  We used zero-mean classical
stationary Gaussian stochastic processes with Gaussian correlations, $\langle
B_i(t)B_j(t')\rangle =\sigma^2 \delta_{ij}e^{-(t-t')^2/\tau_c^2}$, as the
decoherence source.  These are obtained by applying the Gaussian filter to
discrete sets of uncorrelated random numbers drawn from the Gaussian
distribution, and using the standard cubic spline interpolation with the
result.

%%% The core of the program is the pulse sequencer which optimizes the full
%%% list of shaped pulses to be executed in the system and outputs a
%%% sequential list of time intervals with the list of full or partial pulses
%%% to be run in each interval.  As a result, we can efficiently simulate
%%% complex algorithms involing millions of pulses with the number of qubits
%%% only limited by the available memory (the limitation comes from the fact
%%% that full unitary evolution matrix has to be used).

The $[[4,2,2]]$ toric code is a stabilizer code
\cite{Gottesman-Thesis,Nielsen-book} encoding an arbitrary state of
$k=2$ qubits into a $2^2$-dimensional subspace $\mathcal{Q}$ of the
$4$-qubit Hilbert space.  The subspace $\mathcal{Q}$ is the common
$+1$ eigenspace of the two {\em stabilizer generators\/},
$G_x=\sigma^x_1\sigma^x_2\sigma^x_3\sigma^x_4$ and
$G_z=\sigma^z_1\sigma^z_2\sigma^z_3\sigma^z_4$.  We use the following
explicit map (up to normalization)
\begin{eqnarray}
  \hspace{-0.2in}
  \ket{\tilde{0}\tilde{0}}= {\ket{0000}+\ket{1111}},&~ &
  \ket{\tilde{0}\tilde{1}}= {\ket{0011}+\ket{1100}},\nonumber\\
  \hspace{-0.2in}
  \ket{\tilde{1}\tilde{0}}= {\ket{0101}+\ket{1010}},&~ &
  \ket{\tilde{1}\tilde{1}}= {\ket{0110}+\ket{1001}},\quad
  \label{eq:codewords}
\end{eqnarray}
where digits with tildes indicate the states of the logical qubits.
An application of any single-qubit error, i.e., a Pauli operator
$\sigma_i^\mu$, $\mu=x,y,z$, takes the encoded wavefunction to one of
the three orthogonal subspaces, where one or both of the eigenvalues
of $G_x$, $G_z$ (these eigenvalues form the \emph{error syndrome})
equal $-1$.  The code has distance $d=2$ since some two-qubit
errors, e.g., $\sigma_1^z\sigma_2^z$, act within the code space and
cannot be detected.

In the presence of the error Hamiltonian (\ref{eq:noise}), to leading
order in the perturbation, the original wavefunction $\psi_0$ evolves
into a superposition of orthogonal terms $A_0\ket{\psi_0}+A_i^\mu
\sigma_i^\mu\ket{\psi_0}$, where, in general, the coefficients are
operators acting on the bath degrees of freedom, and the state
fidelity is $F\equiv \trb (A_0^\dagger A_0\rho_{\rm B})$, where trace
over bath degrees of freedom with the density matrix $\rho_{\rm B}$ is
taken.  With the same accuracy, $F$ is also the probability that the
measurement returns $G_x=G_z=1$.  For weak perturbation,
and for times $t$ small compared to the bath correlation time
$\tau_c$, the infidelity $1-F$ scales quadratically with $t$; thus
frequent projective measurements of the generators ensure preservation
of the wavefunction with high probability (quantum Zeno
effect\cite{Facchi-Pascazio-2002,Facchi-2002B}).

\begin{figure}[htbp]\center
  \includegraphics[width=0.65\columnwidth]{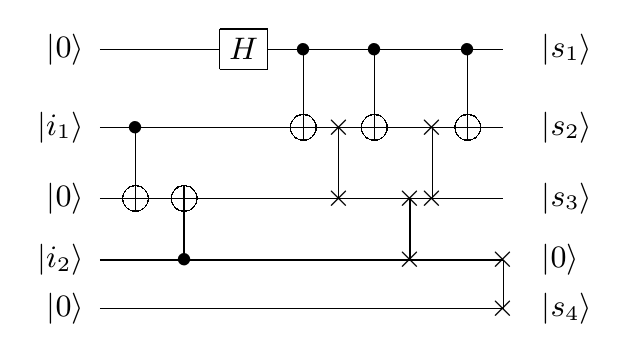}
  \caption{\myfontsize Encoding circuit implemented on a spin chain using
    four exchange gates (implemented with three CNOTs each), five CNOT gates,
    and a Hadamard gate $H$.  Input qubits $i_1$, $i_2$ can be in an arbitrary
    two-qubit state, on the output the circuit returns an equivalent linear
    combination of the states in the code, see Eq.~(\ref{eq:codewords}), using
    qubits $s_j$, $j=1,\ldots,4$, and an ancilla initialized for the
    stabilizer measurement circuit in Fig.~\ref{Meas_cir}.  The decoding is
    done by reversing this circuit.}
  \label{Encod_cir}
\end{figure}

\begin{figure}[htbp]\centering
  \includegraphics[width=0.85\columnwidth]{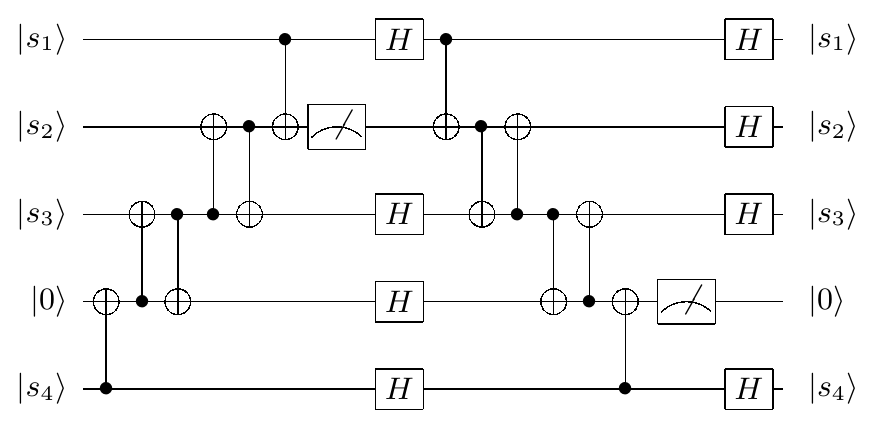}
  \caption{\myfontsize Measurement circuit implemented on a spin chain with a
    shuttling ancilla. The ancilla is first shuttled up for the $G_z$
    measurement and then shuttled down for the $G_x$ measurement (see
    text). The entire circuit is repeated for every Zeno cycle.}
  \label{Meas_cir}
\end{figure}

Using the constructed DD-based gates, we simulated the encoding/decoding and
ancilla-based stabilizer measurement circuits (Figs.~\ref{Encod_cir},
\ref{Meas_cir}), where we use standard quantum circuit
notations\cite{Barenco-1995,Nielsen-book}.  The Hadamard gate was implemented
using the identity $U_{H} =e^{-i\pi/2}\exp({i\frac{\pi}{4}\sigma_y})
\exp({i\frac{\pi}{2}\sigma_x})$, and a projective single-qubit measurement was
implemented simply as an instantaneous projection to the state $\ket0$ of the
ancilla.

\begin{figure}[htbp]
  \raggedleft
  \includegraphics[width=1\columnwidth]{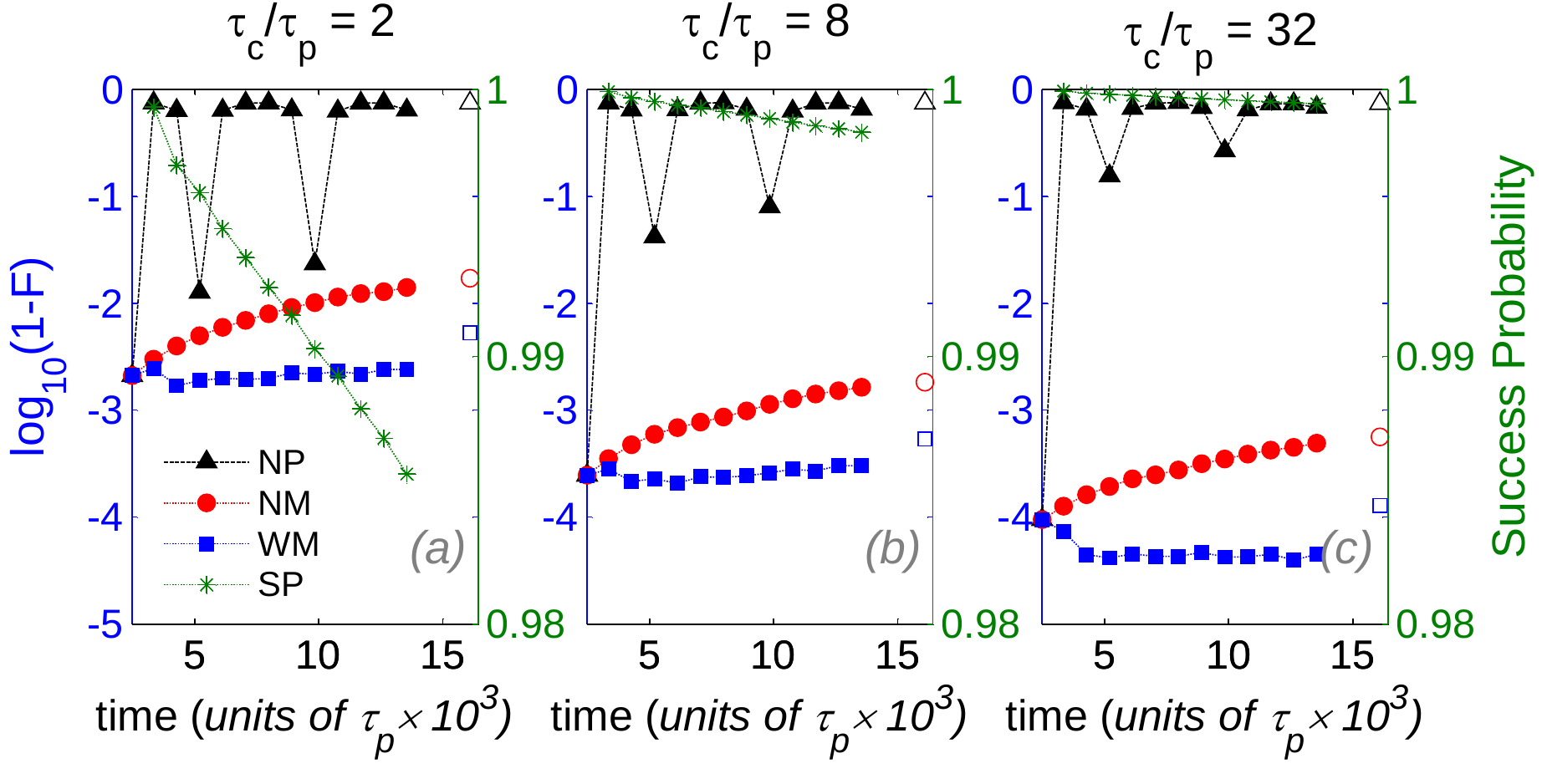}
  \caption{\myfontsize (color online) Infidelities during the Zeno cycle for
    different noise correlation times as shown, with the same noise amplitude
    $\sigma=10^{-3}/\tau_p$. The different curves correspond to cases where no
    pulses are applied (NP), DD pulses are applied but no measurements are
    done (NM), and with the syndrome measurements (WM). Note that the axis for
    the cumulative success probability (SP) of no error is on the right.}
  \label{fig:inf}
\end{figure}

\begin{figure}[htbp]
  \centering
  \includegraphics[width=1\columnwidth]{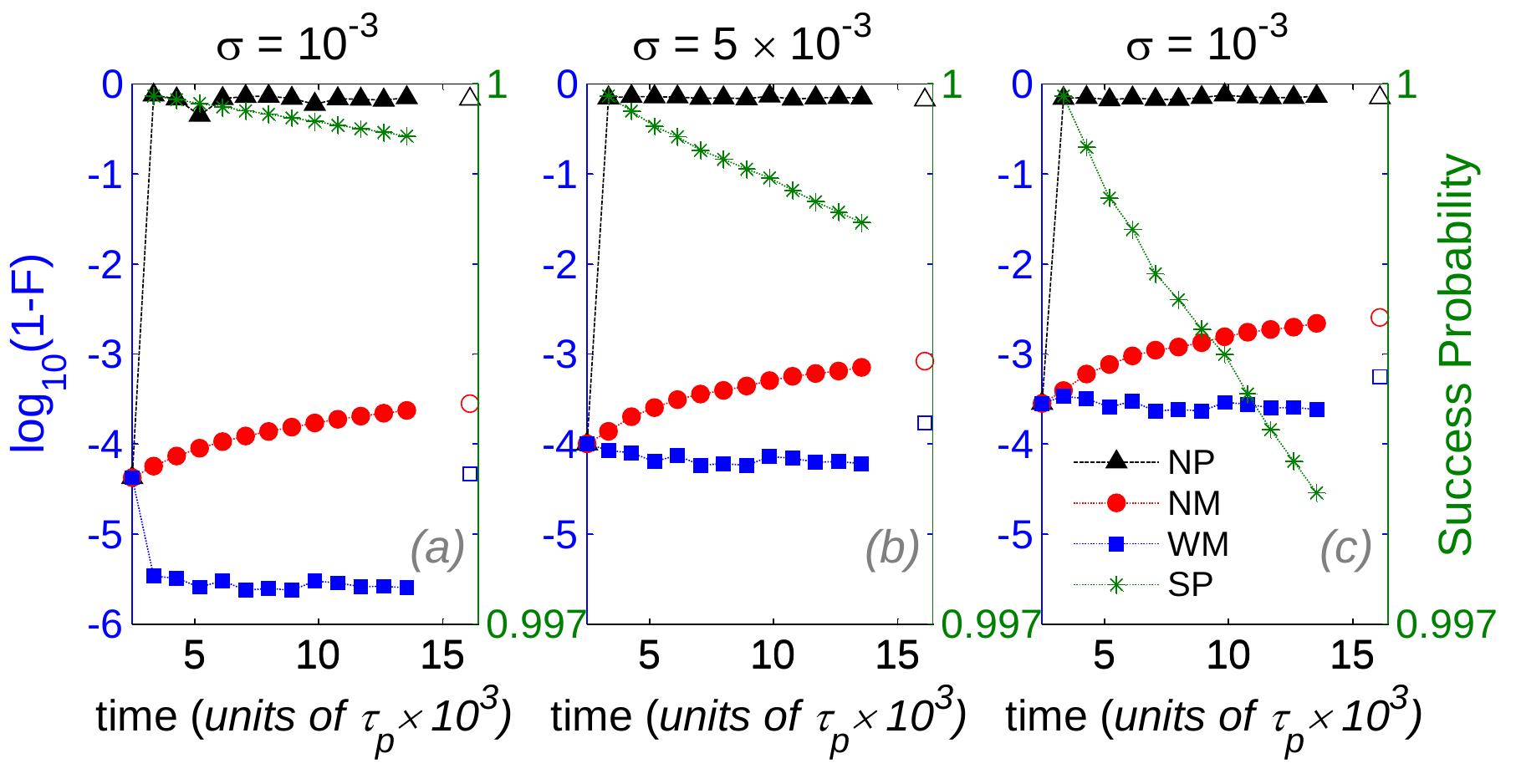}
  \caption{\myfontsize (color online) Same as Fig.~\ref{fig:inf} but with
    fixed noise correlation time $\tau_c=128\tau_p$ with the r.m.s.\ noise
    amplitudes $\sigma$ as indicated.}
  \label{fig:inf_T128}
\end{figure}

Samples of the simulation results are shown in Figs.~\ref{fig:inf},
\ref{fig:inf_T128} which show the time-dependence of average
infidelities and accumulated circuit failure probability, averaged over 20
instances of the classical correlated noise.  The time axis starts at the end
of the encoding circuit, see Fig.~\ref{Encod_cir}.  Closed symbols represent
the computed infidelity after the syndrome measurement (two points per circuit
in Fig.~\ref{Meas_cir}); open symbols at the end of the final decoding
(reverse of the circuit in Fig.~\ref{Encod_cir}).  Blue squares stand for
complete simulation, red circles have the pulses but no projection (dynamical
decoupling is done but no Zeno effect happens, increasing the infidelity in
our simulations by up
to an order of magnitude), and black triangles where neither the decoupling
pulses nor ancilla measurements are applied.

It is clear that the dynamical decoupling and the Zeno effect are both
contributing to improving the fidelity, and both get better with
decreasing noise.  This indicates that the errors contributing to the
infidelity of the constructed gates are not dominated by high-weight
errors which would be undetectable by the code\cite{De-Pryadko-long}.

The infidelity sharply increases with shorter noise correlation
times; this results from the asymmetry of the DCGs, see
Fig.~\ref{fig:pulseY3p}.  We have also constructed\cite{De-Pryadko-long}
 symmetrized DCGs which give second-order
decoupling for arbitrary bath operators $B_i$ when used with the
pulses constructed in Ref.~\onlinecite{Pasini-2008}; we expect such
gates to have a much better accuracy for smaller noise correlation
time, down to the gate duration, $\tau_c\ge 32\tau_p$.

\textsc{In conclusion}, we implemented a universal set of one- and two-qubit
gates for a system with always-on qubit coupling.  The gates are based
on DD techniques and have an added benefit of protection against
low-frequency phase noise.

One application of thus constructed gates is for implementing toric
codes on square lattice, where one sublattice would be used for actual
qubits, and the other sublattice for ancillas.  This way, measurement
of the entire syndrome can be done in just two cycles, each of
four CNOTs in duration, plus some single-qubit gates.  The same
sequences would also work for an arbitrary quantum LDPC code, if the
couplings between the qubits and the ancillas form the corresponding
Tanner graph\cite{Tanner-graph-1981}.  In particular, for
hypergraph-product and related
codes\cite{Tillich-Zemor-2009,Kovalev-Pryadko-2012} one can use the
square lattice layout with additional
connections\cite{Kovalev-Pryadko-FT-2012}.

We are grateful to Kaveh Khodjasteh, Daniel Lidar, and Lorenza Viola for explaining the
working of DCGs.  This work was supported in part by the U.S. Army Research
Office under Grant No.\ W911NF-11-1-0027, and by the NSF under Grant
No.\ 1018935.

%\bibliographystyle{apsrev}
%\bibliography{bibliography,qc_all,lpp,more_qc}

%

\end{document}